\documentclass[%
 reprint,
 amsmath,amssymb,
 aps,
 pra,
]{revtex4-2}

\usepackage{graphicx}
\usepackage{dcolumn}
\usepackage{bm}

\usepackage{braket}
\usepackage{url}
\usepackage{soul}
\usepackage{ textcomp }
\usepackage{placeins}

\usepackage{xcolor}

\newcommand{\Gammatot}{\Gamma} 

\begin{document}

\preprint{APS/123-QED}

\title{Simple analytical model describing the collective nonlinear response of an ensemble of two-level emitters weakly coupled to a waveguide}

\author{Max Schemmer\textsuperscript{\textdagger}}
\altaffiliation[Present address: ]{Istituto Nazionale di Ottica, Consiglio Nazionale delle Ricerche, 50019 Sesto Fiorentino, Italy}
\affiliation{Department of Physics, Humboldt-Universit\"{a}t zu Berlin, 10099 Berlin, Germany}
\author{Martin Cordier}
\thanks{These two authors contributed equally}
\affiliation{Department of Physics, Humboldt-Universit\"{a}t zu Berlin, 10099 Berlin, Germany}
\author{Lucas Pache}
\affiliation{Department of Physics, Humboldt-Universit\"{a}t zu Berlin, 10099 Berlin, Germany}
\author{Philipp Schneeweiss}
\affiliation{Department of Physics, Humboldt-Universit\"{a}t zu Berlin, 10099 Berlin, Germany}
\author{Jürgen Volz}
\affiliation{Department of Physics, Humboldt-Universit\"{a}t zu Berlin, 10099 Berlin, Germany}
\author{Arno Rauschenbeutel}
\affiliation{Department of Physics, Humboldt-Universit\"{a}t zu Berlin, 10099 Berlin, Germany}

\date{\today}


\begin{abstract}
   We model and investigate the collective nonlinear optical response of an ensemble of two-level emitters that are weakly coupled to a single-mode waveguide. Our approach generalizes the insight that photon-photon correlations in the light scattered by a single two-level emitter result from two-photon interference to the case of many emitters. Using our model, we study different configurations for probing the nonlinear response of the ensemble, e.g., through the waveguide or  via external illumination, and derive analytical expressions for the second-order quantum coherence function, $g^{(2)}(\tau)$, as well as for the squeezing spectrum of the output light in the waveguide, $S_\theta(\omega)$. For the transmission of resonant guided light, we recover the same predictions as previously made with more involved theoretical models when analyzing experimental results regarding  $g^{(2)}(\tau)$ (\citet{prasad_correlating_2020}) and $S_\theta(\omega)$ (\citet{hinney_unraveling_2021}). We also study the transmission of light that is detuned from the transition of the two-level emitter, a situation that we recently studied experimentally (\citet{cordier_tailoring_2023}). Our model predictions show  how the collectively enhanced nonlinear response of weakly coupled emitters can be harnessed to generate non-classical states of light using ensembles ranging from a few to many emitters. 
\end{abstract}

\maketitle

\section{Introduction}
Modeling many-body systems is a major aim and challenge in quantum physics. In this context, the description of the scattering of light by an ensemble of quantum emitters is a ubiquitous yet inherently complex problem. Over the past decades, a considerable amount of work has been devoted to the study of collective radiative phenomena. Progress has been made both theoretically \cite{dicke_coherence_1954,gross_superradiance_1982,benedict_super-radiance_1996,bromley_collective_2016,kornovan_collective_2016,asenjo-garcia_exponential_2017,asenjo-garcia_atom-light_2017,kornovan_transport_2017,jones_collectively_2020,sheremet_waveguide_2023}  and experimentally \cite{inouye_superradiant_1999,scheibner_superradiance_2007,van_loo_photon-mediated_2013,araujo_superradiance_2016,corzo_large_2016-1,solano_super-radiance_2017-3,corzo_waveguide-coupled_2019,blaha_beyond_2022,glicenstein_collective_2020,pennetta_observation_2022,pennetta_collective_2022,liedl_collective_2023,ferioli_non-equilibrium_2023,srakaew_subwavelength_2023}. A large part of this research has focused on the single excitation regime or on situations where all emitters are located in a small region of space and couple to the same optical mode, as initially studied by R.~H.~Dicke~\cite{dicke_coherence_1954}. Accounting at the same time for extended ensembles of spatially arranged emitters and many excitations is analytically and even numerically challenging. However, such spatially extended ensembles occur in numerous experimental platforms~\cite{van_loo_photon-mediated_2013,vetsch_optical_2010,solano_super-radiance_2017-3,corzo_waveguide-coupled_2019,araujo_superradiance_2016,ferioli_non-equilibrium_2023}. 

Here, we treat the problem of many two-level emitters that are weakly coupled to a single propagating light mode. Our approach is based on the insight that photon-photon correlations in the light scattered by a single two-level emitter result from two-photon interference~\cite{dalibard_correlation_1983,hanschke_origin_2020, phillips_photon_2020,masters_simultaneous_2023,casalengua_two_2024}. By generalizing this concept to many two-level emitters, we introduce an intuitively accessible model, which allows us to calculate the quantum non-linear response of the ensemble of emitters and to derive analytical expressions for the second-order quantum coherence function and the squeezing spectrum of the output light. In order to check the validity of our model, we compare its predictions to those obtained in~\cite{mahmoodian_strongly_2018-1}, where it has been shown theoretically that continuous wave input laser light can be transformed into perfectly anti-bunched or strongly bunched light depending on the number of coupled emitters, and find perfect agreement. Compared to previous theoretical descriptions, the simplicity of our approach results in a significant computational advantage and provides a much more intuitive picture. 
Furthermore, it can account for different probing configurations, e.g., through the waveguide or via external illumination, and different spatial arrangements of the emitters. 
Using our model predictions, we exemplify how to exploit the collective nonlinear response of an ensemble of emitters for, e.g., generating quantum light with tailored photon statistics or for realizing a narrowband source of squeezed light. 
Finally, while the transformation of laser light into antibunched light with a single quantum emitter is well-established~\cite{kimble_photon_1977} and was recently also shown to be possible with large ensembles of weakly coupled emitters~\cite{prasad_correlating_2020}, here, we anticipate that this is also possible with a small number (for example $N=3$) of suitably arranged emitters by illuminating them simultaneously form the outside and through the waveguide.
This makes our model useful and attractive to a broad community working with different platforms, ranging from solid-state approaches, where the number of spectrally identical emitters is typically limited to a few~\cite{lodahl_interfacing_2015}, to quantum nanophotonics with trapped cold atoms, where the number of spectrally identical emitters can reach thousands.

{This article is organized as follows: In Sec.~\ref{sec_definitions}, we introduce the fundamental quantity underlying the physics of this article, the two-photon wavefunction $\psi(\tau)$. We then decompose the two-photon wavefunction into its coherent and incoherent components and work out its relation to the second-order quantum correlation $g^{(2)}(\tau)$ function as well as to the quadrature squeezing spectrum $S_\theta(\omega)$. In Sec.~\ref{sec_single_emitter} we explain the significance of the coherent and incoherent component with the pedagogical example of resonance fluorescence from a single emitter where the resulting antibunched light can be understood as a destructive interference of both components~\cite{dalibard_correlation_1983,hanschke_origin_2020, phillips_photon_2020,zubizarreta_casalengua_conventional_2020,masters_simultaneous_2023,casalengua_two_2024}. Furthermore, we discuss different illumination configurations for a single emitter coupled to a guided mode (Sec.~\ref{sec_single_emitter_external} - Sec.~\ref{sec_single_combined}). In Sec.~\ref{sec_many_emitters}, we introduce our model which extends the interference concept from a single emitter to an ensemble of emitters coupled to the same waveguide mode. We then apply our model to case studies  of emitters coupled to a waveguide under different excitation geometries. First, we recover with our model known results for emitters resonantly excited through the waveguide~\cite{mahmoodian_strongly_2018-1} (Sec.~\ref{sec:Transmission}), and then extend it to new geometries such as illumination under the Bragg (Sec.~\ref{sec_Bragg}) and anti-Bragg angle (Sec.~\ref{sec_anti_Bragg}), as well as a  combined illumination from outside the waveguide and through the waveguide (Sec.~\ref{sec_combined}).}

\section{Definitions}\label{sec_definitions}
We seek to model the transformation of a weak classical continuous-wave monochromatic light field upon interaction with an ensemble of $N$ two-level emitters. We assume the latter to be weakly coupled to the same propagating optical mode, which is defined, e.g., by a single-mode optical waveguide~\footnote{While, here, we limit the discussion to waveguide coupled emitters, the approach can be readily extended to emitters in free-space, such as in~\cite{masters_simultaneous_2023}.}. This situation can be experimentally realized with various emitters, such as laser cooled atoms or molecules in organic host material coupled to nanofibers~\cite{vetsch_optical_2010,liebermeister_tapered_2014,skoff_optical-nanofiber-based_2018}, 
or with quantum dots, color centers, or excitons coupled to photonic waveguides~\cite{akimov_generation_2007,lund-hansen_experimental_2008,friedler_solid-state_2009,sapienza_cavity_2010,faez_coherent_2014,lodahl_interfacing_2015,sipahigil_single-photon_2016,aharonovich_solid-state_2016,lombardi_photostable_2018,boissier_coherent_2021,turschmann_coherent_2019,shreiner_electrically_2022}. In the microwave regime, this can be also realized with superconducting qubits coupled to a transmission line~\cite{astafiev_resonance_2010,hoi_microwave_2013}. While realizations with laser cooled atoms allow one to couple large numbers of weakly-coupled identical emitters~\cite{gouraud_demonstration_2015,corzo_large_2016-1,pennetta_collective_2022,liedl_collective_2023}, solid state emitters allow up-to-now only the coupling of few identical emitters, with, however, much larger coupling strength~\cite{mlynek_observation_2014,mirhosseini_cavity_2019,chu_independent_2023,duquennoy_real-time_2022}. The model we present in this manuscript allows us to account for both, few and many weakly coupled identical emitters.   

\begin{figure}[bt]
\centering
\includegraphics[width=0.8\linewidth]{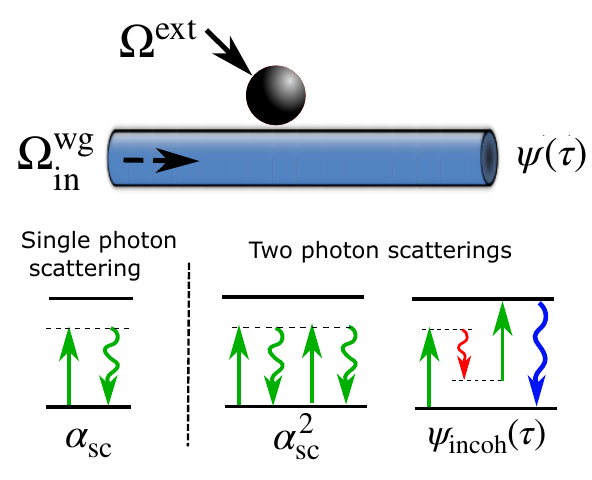}
\caption{Top: A single emitter coupled to a waveguide mode. The emitter can be driven externally ($\Omega^\mathrm{ext}$) and / or through the waveguide ($\Omega^\mathrm{wg}_\mathrm{in}$). Bottom: Possible one and two photons scattering processes with the coherent single photon scattering amplitude $\alpha_{sc}$, the coherent two-photon scattering amplitude $\alpha_{sc}^2$ and the incoherent two-photon scattering amplitude $\psi_\mathrm{incoh}(\tau)$ which is given in Eq.~\eqref{eq_phi_1_tau}.}
\label{fig_singleatom}
\end{figure}

The aim of our model is to be able to describe the output quantum state of light in the waveguide mode when the ensemble of emitters is weakly driven with continuous-wave monochromatic laser light with frequency $\omega_L$, free-space wavelength $\lambda = 2\pi c/\omega_L$, detuning $\Delta = \omega_L - \omega_0$ from the atomic transition frequency $\omega_0$, and the resonant Rabi frequency $\Omega$. The latter is complex, $\Omega = |\Omega| e^{i \phi}$, where the phase $\phi$ is defined by the phase of the driving field in the absence of emitters. 
In our model, we explore different illumination geometries for emitters coupled to a waveguide. The emitters can be illuminated either through the waveguide, externally, or simultaneously through both pathways, as illustrated in Fig.~\ref{fig_singleatom}.
Each of the $N$ emitters with the population decay rate of the excited state $\Gamma$ is coupled to the forward-direction of a waveguide with a coupling efficiency  $\beta = \Gamma_\text{wg}/\Gammatot$ where $\Gamma_\text{wg}$ is the emission rate of the emitters into the forward-direction of the waveguide. In our model, we focus on the weak drive regime $\Omega \ll \Gamma$, where the driving field can be approximated by a truncated coherent state with at most two photons 
\begin{equation}
    \ket{\alpha_\mathrm{in}} = e^{-|\alpha_\mathrm{in}|^2/2} \left(\ket{0} + \alpha_\mathrm{in}\ket{1} + \frac{\alpha_\mathrm{in}^2}{\sqrt{2}}\ket{2}\right)~.
\end{equation} Using this truncation allows us to describe the output state after the interaction with the ensemble of emitters as~\cite{mahmoodian_strongly_2018-1}: 
\begin{equation}
        \ket{\mathrm{out}} = \ket{\mathrm{out}}_\mathrm{1\rightarrow1} + \ket{\mathrm{out}}_\mathrm{2\rightarrow2}+\ket{\mathrm{out}}_\mathrm{2\rightarrow1}
    \label{eq:states}
\end{equation}
where $\ket{\mathrm{out}}_\mathrm{X\rightarrow Y}$ describes a state of the combined system of guided and unguided modes with  $Y$ of the $X$ input photons scattered into the forward direction of the waveguide. Correspondingly, for the component $\ket{\mathrm{out}}_\mathrm{2\rightarrow1}$, one of the two input photons has not been scattered into the waveguide mode but into unguided modes~\footnote{{Note that, when tracing over the unguided modes, a density matrix description for the output state would, in principle, be required. However, since in this work the state $\ket{\mathrm{out}}_{\mathrm{2\rightarrow1}}$ does not contribute, the description in terms of pure states is sufficient.~\cite{mahmoodian_strongly_2018-1}.}}. Note that we omit states with $Y=0$ as they do not contribute to the observables of interest (second-order correlation function $g^{(2)}$ and squeezing spectrum $S_\theta(\omega)$)~\cite{mahmoodian_strongly_2018-1,hinney_unraveling_2021}.  

The origin of the different components of the output state will become clearer when deriving them for distinct scattering processes. 
Each of the emitters can either coherently or incoherently scatter the incoming photons~\footnote{Here, coherent and incoherent scatterign refers ti the first-order coherence properties of the two components, as commonly used in literature~\cite{steck_quantum_2007}.} as illustrated in Fig.~\ref{fig_singleatom}.  As coherent scattering only involves a modification of the amplitude and phase, it is straightforward to account for. Calculating the impact of incoherent scattering on the output state turns out to be more involved. Incoherent scattering involves the scattering of two or more photons and, therefore, in our truncated Hilbert space, it only affects states with $X=2$.
In the weak drive regime, where single-photon scattering dominates, the total output power is predominately governed by power of the coherent component of the output state, which scales linearly with the power of the driving field, $P_\mathrm{coh} \propto |\Omega|^2$. By contrast, the power of the incoherent component of the output state scales quadratically with the power of the driving field, $P_\mathrm{incoh} \propto |\Omega|^4$. However, while the incoherent component thus only contributes a negligible fraction to the total output power in the weak drive regime, it plays a crucial role in observables such as the normalized second-order correlation function, $g^{(2)}(\tau)$, or the squeezing spectrum, $S_\theta(\omega)$. These observables can be both inferred from the \textit{temporal two-photon wavefunction}, $\psi(t_1,t_2)$, which in the stationary regime depends only on the relative time $\tau = t_2 - t_1$, and is associated with $\ket{\mathrm{out}}$ via the relation:
\begin{align}
    \ket{\mathrm{out}}_{2\rightarrow2} &= A \iint \psi(t_2 - t_1)  \, \hat{a}^\dag{}(t_1)\hat{a}^\dag{}(t_2)    dt_1 dt_2 \ket{0}
    \label{eq: wavefunction}
\end{align}
Similarly, $\ket{\mathrm{out}}$ can be represented in frequency space by $\psi(\omega)$ as 
\begin{align}
    \ket{\mathrm{out}}_{2\rightarrow2} &= \frac{A}{2\pi} \iint \delta(\omega + \omega') \psi(\omega)  \, \hat{a}^\dag{}(\omega)\hat{a}^\dag{}(\omega')  d\omega d\omega'  \ket{0}
\end{align}
where $\psi(\omega) = \int d \tau \psi(\tau) e^{i \omega \tau}$.
Here, the normalization $A$ is chosen such that the coincidence count rate is $G^{(2)}(\tau) = \langle \hat{a}^\dag(0)\hat{a}^\dag(\tau) \hat{a}(0)\hat{a}(\tau) \rangle  = |\psi(\tau)|^2 $ and $\hat{a}^{\dagger}(t)$  is the standard bosonic creation operator together with its frequency representation $\hat{a}^{\dagger}(\omega)$. 

The two-photon wavefunction can be decomposed into the \textit{coherent} and \textit{incoherent two-photon components} $\psi(\tau)  = \psi_\mathrm{coh} + \psi_\mathrm{incoh}(\tau)$, cf.~Fig.~\ref{fig_singleatom}, where the coherent component contains the coherently scattered component by the emitters, as well as - depending on the geometry - the residual of the driving field. For a continuous wave input the coherent component is time independent, while the incoherent two-photon component depends on the relative time $\tau$. In leading order of the driving field ($\Omega^2$), the normalized second-order correlation function is then given by~\cite{mahmoodian_strongly_2018-1} 
\begin{align}
    g^{(2)}(\tau) &= \frac{|\psi(\tau)|^2}{|\alpha_\mathrm{out}|^4}=\frac{|\psi_\mathrm{coh} + \psi_\mathrm{incoh}(\tau)|^2}{|\alpha_\mathrm{out}|^4}
    \label{eq:g2}
\end{align}
where $\alpha_\mathrm{out}$ is the coherent single photon component.
From Eq.~\eqref{eq:g2}, one can see that interference between the coherent and incoherent component of the two-photon wavefunction will impact the value of $g^{(2)}(\tau)$ \cite{dalibard_correlation_1983,zubizarreta_casalengua_conventional_2020}.

Another signature of the quantum correlations that result from the interaction with many emitters is a reduction of the quantum fluctuations for one quadrature component, i.e., quadrature squeezing~\cite{heidmann_squeezing_1985,hinney_unraveling_2021}. The latter can be quantified by means of the normally ordered squeezing spectrum,
\begin{equation}
S_{\theta}(\omega)=\int_{-\infty}^{\infty}\left\langle: \Delta \hat{X}_{\theta}(\tau) \Delta \hat{X}_{\theta}(0):\right\rangle e^{i \omega \tau} d \tau~.
\end{equation}
Here, we introduced the generalized quadrature operator $\hat{X}_{\theta}(t) = \frac{1}{2}\left[\hat{a}_t e^{i \theta}+\hat{a}^{\dagger}_t e^{-i \theta}\right]$ with $\Delta \hat{X}_{\theta}(t) = \hat{X}_{\theta}(t) - \braket{\hat{X}_{\theta}(t)}$ where $:...:$ denotes normal ordering.  In the weak drive regime ($\Omega \ll \Gamma$), the squeezing spectrum is then approximately given by~\cite{hinney_unraveling_2021,kusmierek_higher-order_2023}: 
\begin{equation}
S_{\theta}(\omega)= - \frac{1}{2}\left|\psi_\mathrm{incoh}(\omega)\right| \cos \left[2 \theta+\varphi(\omega)\right] 
\label{eq_squeezing}
\end{equation}
 with $\varphi(\omega) = \text{Arg}\lbrace\psi_\mathrm{incoh}(\omega)\rbrace$ where we introduced the Fourier transform $\psi_\mathrm{incoh}(\omega) = \int \psi_\mathrm{incoh}(\tau) e^{i\omega\tau} d\omega$. Equation~\eqref{eq_squeezing} links the experimentally accessible squeezing spectrum $S_\theta(\omega)$ to the incoherent component of the two-photon wavefunction $\psi_\mathrm{incoh}(\omega)$, as demonstrated in~\cite{hinney_unraveling_2021}.

In the following, we will describe how to calculate $g^{(2)}(\tau)$ as well as $S_{\theta}(\omega)$ for different spatial arrangements of emitters and configurations of illumination.

\section{Single emitter}\label{sec_single_emitter}
Let us first consider the case of a single emitter coupled to a single mode waveguide as sketched in Fig.~\ref{fig_singleatom}. The emitter is driven by a field that is either external or guided and has a resonant Rabi frequency $\Omega$ with $|\Omega|\ll \Gammatot$ and a detuning $\Delta$. Correspondingly, in leading order the atom scatters photons coherently and incoherently with a total rate $R = |\Omega|^2/\Gamma (1 +(2\Delta/\Gamma)^2)$.
The scattered photons are coupled into the forward direction of the guided mode with probability $\beta$. The amplitude of coherently scattering a single photon into the forward direction of the guided mode is then
\begin{align}
\alpha_\mathrm{sc} = - \frac{\Omega}{2\sqrt{\beta \Gamma}}\ g_\Delta
\label{eq:alphasc}
\end{align}
where we introduce a power-independent photon generation coefficient
\begin{equation}
g_{\Delta} = \frac{2\beta }{1-2i\Delta/\Gamma}.
\end{equation}
The minus sign in Eq.~\eqref{eq:alphasc} stems from the $\pi$ phase shift between the scattered light field and the driving field for on-resonant excitation~\footnote{The $\pi$-phase shift on resonance is composed from two contribution: First, a $\pi/2$ from coupling into the waveguide (or the Gouy-phase in free-space) and, second, the dipole is oscillating with $\pi/2$ out-of-phase at resonance with respect to the excitation laser.}.
The amplitude of generating \textit{two} coherently scattered photons in the waveguide is independent of their time difference $\tau$ and given by  $\alpha_\mathrm{sc}^2$.
The amplitude of the incoherently scattered two-photon component in forward direction is given by~\cite{mahmoodian_strongly_2018-1, le_jeannic_experimental_2021}: 
\begin{equation}
\psi_\mathrm{incoh}(\tau) = - \alpha_\mathrm{sc}^2 \, e^{-\Gammatot |\tau|/2} e^{i \Delta |\tau|}. \label{eq_phi_1_tau}
\end{equation}
For $\tau = 0$, the amplitude of this nonlinear scattering process is equal in magnitude to the coherent amplitude of scattering two photons into the waveguide $\alpha_\mathrm{sc}^2$. For finite $\tau = t_2 - t_1$, it exponentially decays with $|\tau|$. 
The photon-pairs created from an incoherent scattering are thus correlated in time with a typical separation $\tau_\mathrm{exc} = 1/\Gammatot$ -- the lifetime of the excited state. This is in contrast with $\psi_\mathrm{coh}$ that describes two temporally uncorrelated photons. 

\begin{figure}[tb]
\centering
\includegraphics[width=0.9\linewidth]{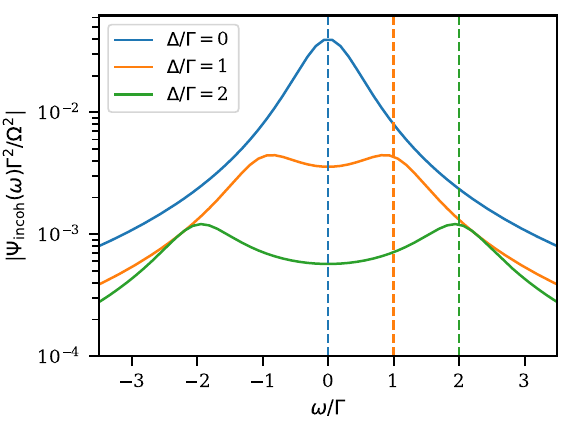}
\caption{Frequency spectrum: Absolute value of the incoherent two-photon amplitude $\psi_\mathrm{incoh}(\omega)$ for a single emitter and for three different detunings $\Delta$. The spectra are plotted in the excitation laser frame, such that $\omega=0$ corresponds to the laser frequency. The dashed lines indicate the respective detunings and thus coincide with the resonance frequency of the atom. For sufficiently large detunings, the spectrum of the entangled photon pairs is peaked around $\pm \Delta$.}
\label{fig_off_res_phi_1}
\end{figure}

In the frequency domain, the incoherent scattering can be understood as a degenerate four-wave mixing process where two laser photons are converted into a pair of red- and blue-detuned photons of frequency $\omega_L \pm \omega$ where $\omega$ is the detuning from the laser frequency $\omega_L$. The spectrum of the incoherent component is given by~\cite{le_jeannic_experimental_2021}:
\begin{align}
    \psi_\mathrm{incoh}(\omega)  &=  -\frac{\Omega^2}{2 \beta^2 \Gammatot^2} g_\Delta g_{\Delta+\omega} g_{\Delta-\omega}~.
    \label{eq:phi}
\end{align}
For a resonant laser excitation, $\Delta=0$, this spectrum has a Lorentzian shape of full width at half maximum (FWHM) $\Gammatot$ as shown in Fig.~\ref{fig_off_res_phi_1}. As a consequence of energy conservation, the generated red- and blue-detuned photons are time-frequency entangled~\cite{ shen_strongly_2007-2,hinney_unraveling_2021,le_jeannic_experimental_2021}. For off-resonant excitation,  $\Delta \gtrsim \Gammatot$, the spectrum consists of two Lorentzians of width $\Gammatot$ with one being centered on the atomic resonance at $\omega_0 $ and one detuned by $2\Delta$ from the atomic-resonance, i.e., at $\pm\Delta$ with respect to the driving frequency, see Fig.~\ref{fig_off_res_phi_1}.

\subsection{Single emitter, external illumination}\label{sec_single_emitter_external} 

In the case of external illumination of a single emitter with Rabi frequency $\Omega^\mathrm{ext}$, the coherent component of the two-photon wavefunction of the output guided field solely stems from the light scattered by the emitter, such that~\footnote{Note that our definition follows~\cite{mahmoodian_strongly_2018-1,kusmierek_higher-order_2023} which differs from~\cite{cordier_tailoring_2023,masters_simultaneous_2023} by a factor $1/2$.}: 
\begin{equation}
    \psi_\mathrm{coh} = \alpha_\mathrm{sc}^2~.
    \label{eq:psicoh_1emitter}
\end{equation}

\begin{center}
\begin{figure*}[bt]
    \centering
    \includegraphics[width=0.85\linewidth]{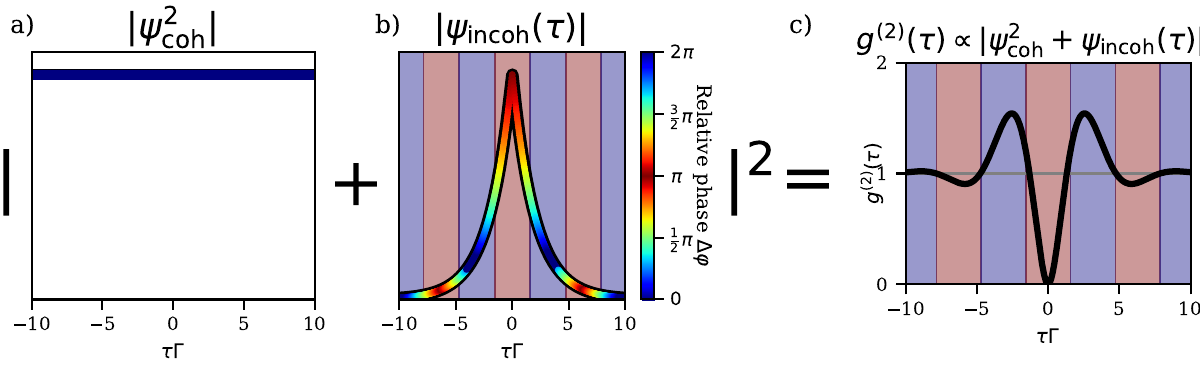}
    \caption{Scattered two-photon wavefunctions from a single emitter at detuning  $\Delta/\Gammatot = 1$ undder external illumination: a) The coherent component $\psi_\mathrm{coh}$ is time-independent and has a constant phase. b) The incoherent component $\psi_\mathrm{incoh}(\tau)$ is a localized function, and the colors correspond to the phase difference to the coherent component $\Delta\varphi = \mathrm{Arg}\lbrace \psi_\mathrm{incoh}(\tau)\rbrace - \mathrm{Arg}\lbrace \psi_\mathrm{coh}\rbrace$. At $\tau=0$, the phase difference is $\pi$ such that in c) antibunching occurs. The second-order correlation function $g^{(2)}(\tau)$ is obtained from a) \& b) by taking into account the complex phase. The constructive/destructive nature of the interference is indicated by the background shading in b) \& c) such that at $\tau =0$ a fully destructive interference occurs. The interference pattern as function of $\tau$ then leads leads signatures of Rabi-oscillation in $g^{(2)}(\tau)$.}
    \label{fig_single_emitter}
\end{figure*}
\end{center}

Combining Eq.~\eqref{eq_phi_1_tau} and~\eqref{eq:psicoh_1emitter} , the total temporal two-photon wavefunction is then:
\begin{equation}
    \psi(\tau) = \alpha_\mathrm{sc}^2 (1 - e^{-\Gammatot |\tau|/2} \, e^{i \Delta |\tau|})~.
    \label{eq:phi1}
\end{equation}
Note that, at zero time delay, both components interfere fully destructively such that $\psi(\tau=0)=0$, as expected for a single quantum emitter. This highlights the fact that, in the low saturation regime, the antibunching of a single two-level emitter results from the destructive interference between coherently and incoherently scattered photons~\cite{dalibard_correlation_1983}.  
This interference is illustrated in Fig.~\ref{fig_single_emitter} and was recently studied in several experiments~\cite{hanschke_origin_2020, phillips_photon_2020,ng_observation_2022,masters_simultaneous_2023} which showed that the photon statistics can be tuned by modifying the balance of coherent and incoherent light, e.g. via spectral filtering of either the coherent or incoherent component from a single two-level emitter~\cite{casalengua_two-photon_2023,casalengua_two_2024}.

\subsection{Single emitter, illumination through the waveguide}
\label{sec_single_emitter_waveguide}
Now let us consider the case where the emitter is driven through the waveguide with Rabi frequency $\Omega_\mathrm{in}^\mathrm{wg}$, see Fig.~\ref{fig_singleatom}, with the corresponding amplitude of the input coherent state $\alpha_\mathrm{in} = \Omega^\mathrm{wg}_\mathrm{in}/(2\sqrt{\beta \Gamma})$. 
The coherent component at the waveguide output $\alpha_\mathrm{out}$ now consists of the coherently scattered light $\alpha_\mathrm{sc}$ that interferes with the coherent field that is already present inside the waveguide $\alpha_\mathrm{in}$: 
\begin{equation}
    \alpha_\mathrm{out} = \alpha_\mathrm{in} + \alpha_\mathrm{sc}= t_\Delta \alpha_\mathrm{in} 
    \label{eq:alphaout}
\end{equation}
where we define $t_\Delta = 1 - g_\Delta$ as the complex-valued single atom transmission coefficient. It follows that the coherent component of the two-photon wavefunction is:
\begin{equation}
    \psi_\mathrm{coh} = t^2_\Delta \alpha^2_\mathrm{in}~.
\end{equation}
Adding the coherent and incoherent component of the two-photon wavefunction, we obtain the total two-photon wavefunction:
\begin{align}
    \psi(\tau) &= t^2_\Delta\alpha_\mathrm{in}^2 - \alpha_\mathrm{sc}^2  e^{-\Gammatot |\tau|/2} e^{i \Delta |\tau|} \nonumber \\
    &= \alpha_\mathrm{in}^2 \big( t^2_\Delta- \frac{4\beta^2}{(1-2i\Delta/\Gamma)^2}e^{-\Gammatot |\tau|/2} e^{i \Delta |\tau|}  \big)~.
\end{align}
In the limit of $\beta \ll 1$, the scattered field of a single atom is much smaller than the input field, $\alpha_\mathrm{sc} \ll \alpha_\mathrm{in}$, such that the two-photon wavefunction at the output is nearly identical to the one of the input, $\psi(\tau) \approx \alpha_\mathrm{in}^2$. Thus, a single weakly coupled emitter has a negligible effect on the photon statistics of the light transmitted in the waveguide. 
In contrast, as discussed in Sec.~\ref{sec:Transmission}, the interaction of $N$ weakly coupled emitters, can strongly suppress or enhance the two-photon wavefunction at zero time delay $\psi(\tau =0)$.

\subsection{Single emitter, combined illumination}\label{sec_single_combined}
Let us now consider the case where the emitter is driven by both an external field and a guided field with Rabi frequencies $\Omega^\mathrm{ext}$ and $\Omega_\mathrm{in}^\mathrm{wg}$, respectively. Here and in the following, when considering combined illumination, for simplicity we assume the two fields in phase with identical polarizations and detunings, such that the Rabi frequencies sum up to the total Rabi frequency seen by the atom $\Omega = \Omega^\mathrm{ext} + \Omega_\mathrm{in}^\mathrm{wg}$. 
The combined scattering rate is then
\begin{align}
    \bar{\alpha}_\mathrm{in}^\mathrm{sc} &= -(\Omega_\mathrm{in}^\mathrm{
wg} +  \Omega^\mathrm{
ext}) \frac{ (\beta\Gamma)^{1/2}}{\Gamma -2i\Delta}\\
&= -\Omega_\mathrm{in}^\mathrm{wg} \frac{(\beta'_\mathrm{in}\Gamma)^{1/2}}{\Gamma -2i\Delta}\label{eq:alphasc_combined}
\end{align}
where we introduced the effective coupling constant $\beta'  = \beta \left( 1 + \Omega^\mathrm{ext}/\Omega_\mathrm{in}^\mathrm{wg}\right) $ which depends on the ratio of the driving field $\Omega_\mathrm{in}^\mathrm{wg}$ and the field resulting from the external Bragg beam $\Omega^\mathrm{ext}$. 
In analogy to Sec.~\ref{sec_single_emitter_waveguide}, one can introduce an effective transmission $t'_\Delta$ and normalized scattering coefficient $g'_\Delta$
\begin{equation}
    {t}'_\Delta = 1 -  \frac{2\beta'}{1 - 2i\Delta/\Gammatot}  =  1  - g'_\Delta
\end{equation}
and accordingly, the coherent component of the two-photon wavefunction is given by:
\begin{align}
    \psi_\mathrm{coh} &= \alpha_\mathrm{out}^2 \nonumber\\
    & = \left(\alpha_\mathrm{in} + \alpha_\mathrm{sc}\right)^2 = \alpha_\mathrm{in}^2  {t'}_\Delta^2
    \label{eq:cohcombined}
\end{align}
The incoherent component of two-photon wavefunction $\psi_\mathrm{incoh}$ is given by the same expression as in Eq.~\eqref{eq_phi_1_tau} where $\alpha_\mathrm{sc}$ is given by Eq.~\eqref{eq:alphasc_combined}. 

Finally, adding the expressions in Eq.~\eqref{eq:cohcombined} and Eq.~\eqref{eq_phi_1_tau}, we obtain the two photon wavefunction in the case of combined illumination of a single emitter:
\begin{equation}
    \psi(\tau) = \alpha_\mathrm{in}^2 \big( t'^2_\Delta- \frac{4\beta'^2}{(1-2i\Delta/\Gamma)^2}e^{-\Gammatot |\tau|/2} e^{i \Delta |\tau|}  \big)~.
\end{equation}
Note that in general $\beta'$ is a complex number where $\mathrm{Arg}\lbrace \beta\rbrace$ is given by the relative phase  between $\Omega^\mathrm{ext}$ and 
$\Omega^\mathrm{wg}$. In the following we restrict our discussions to the case where $\Omega^\mathrm{ext}$ and $\Omega^\mathrm{wg}$ are perfectly in phase such that $\beta'$ is real and $\beta'\geq 1$.

In summary, under a combined illumination, a single emitter effectively behaves, as if illuminated through the waveguide with an effectively increased coupling constant $\beta'$ when the fields are in phase, as discussed in details in~\cite{goncalves_unconventional_2021-1}.

\section{Many emitters}\label{sec_many_emitters}
Let us now consider $N$ emitters weakly-coupled to a waveguide ($\beta \ll 1$) where the distance between emitters is large enough to neglect direct dipole-dipole coupling via the near-field between emitters~\footnote{Near-field effects can typically be neglected when the atomic distances $d$ are such that $d \geq \lambda/2$~\cite{asenjo-garcia_atom-light_2017}.}. Here, we consider the case of unidirectional coupling, i.e., we assume the emission rate into the backward direction of the waveguide to be zero. This reduces the complexity of the problem and can be realized to a good approximation with, e.g., cold trapped atoms coupled to nanofibers~\cite{meng_imaging_2020}. Moreover, in the weak coupling regime, many results obtained in this paper for unidirectional coupling are also valid for bidirectional coupling, see Appendix~\ref{sec_SM_unidirectional} for a more detailed discussion. 

The coherent component at the waveguide output now consists of the sum of the input guided field and the coherent scattering amplitudes of all emitters, $\alpha_n^\mathrm{sc}$, where the index $n = 1,~2,~3,\ldots$ numbers the atoms:
\begin{equation}
    \alpha_\mathrm{out} = \alpha_\mathrm{in}+ \sum_{n=1}^N \alpha_{\mathrm{sc},n}~.
    \label{eq_alpha_wvd} 
\end{equation}
Here, the coherent scattering amplitudes $\alpha_{n}^{\textrm{sc}}$  are computed according to Eq.~\eqref{eq:alphasc} using the local Rabi frequencies $\Omega_n$ seen by the $n$-th atoms, which depend on the illumination configuration, see below. As previously, the coherent two-photon component is then given by $\psi_\mathrm{coh} = \alpha_\mathrm{out}^2$.

For deriving the incoherent two-photon component at the waveguide output, we resort to the frequency domain and find: 
\begin{equation}
    \psi_\mathrm{incoh}(\omega) = \sum_{n = 1}^N  \psi_\mathrm{incoh}^{(n)}(\omega) \left(t_{\Delta + \omega} t_{\Delta - \omega}\right)^{(N-n)}.\label{Phi_N_delta}
\end{equation}
where $\psi_\mathrm{incoh}^{(n)}(\omega)$ is the frequency-domain representation of the incoherent component of the two-photon amplitude of the $n$-th atom in the chain. We calculate it according to Eq.~\eqref{eq:phi} using the local Rabi frequency $\Omega_n$, thereby assuming that the incoherent scattering process is only driven by the coherent component of the incident field. The product of amplitude transmission coefficients, $t_{\Delta + \omega} t_{\Delta - \omega}$, gives the amplitude transmission coefficient of a pair of incoherently scattered blue- and red-detuned photons past one atom. Note that, here, we neglect higher order nonlinear interaction where incoherently scattered photon pairs experience a second non-linear scattering. Neglecting these higher-order incoherent scattering processes is a good approximation in the regime of weak coupling ($\beta \ll 1$), see Sec.~\ref{sec:Transmission} where we explicitly compare our result with~\cite{mahmoodian_strongly_2018-1}.

The output total two-photon amplitude in the time domain is thus given by $\psi(\tau) = \psi_\mathrm{coh} + \psi_\mathrm{incoh}(\tau)$. In the following, we apply this formalism to derive quantitative predictions for four experimentally interesting situations. 


\begin{figure*}[hbtp]
\centering
\includegraphics[width=0.7\linewidth]{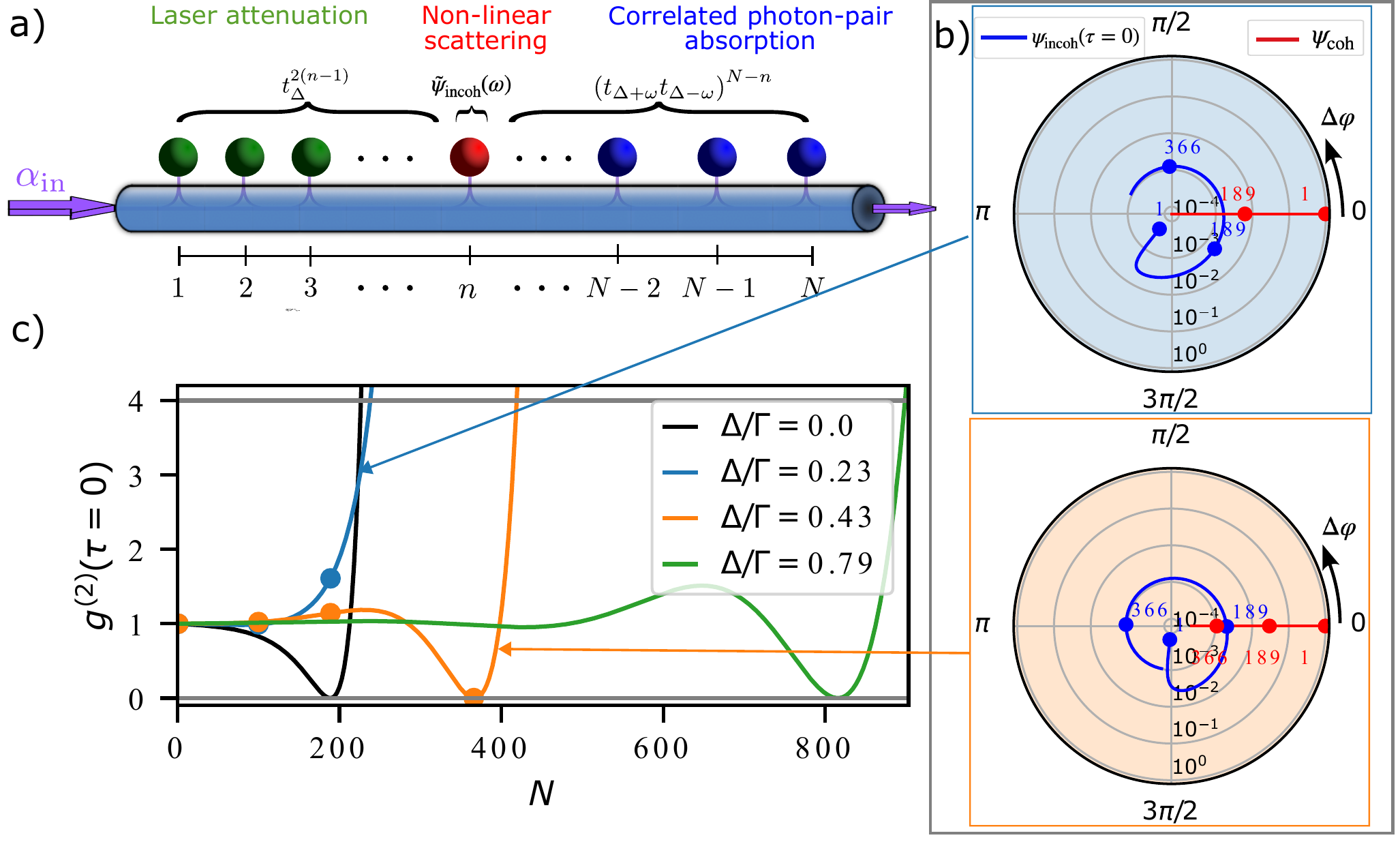}
\caption{a) Contribution of the $n$-th atom to the incoherent two-photon component of the transmitted light. The first $n-1$ emitters attenuate the input light which is then incoherently scattered by the $n$-th emitter. Then, the following $N-n$ emitters act as a dispersive and absorptive medium for this incoherent two-photon component. b) The evolution of the coherent amplitude $\psi_\mathrm{coh}$ and incoherent photon amplitude $\psi_\mathrm{incoh}(\tau = 0)$ at zero time delay as a function of atom number for two different detunings $\Delta$. The amplitudes are normalized to $\alpha_\mathrm{in}$. c) Second-order correlation function at zero time delay, $g^{(2)}(\tau=0)$, for different detunings. Plots are shown for $\beta = 0.007$ as in~\cite{cordier_tailoring_2023}.   
}
\label{fig_sketch_N_atoms}
\end{figure*}

\subsection{N emitters, illumination through the waveguide} \label{sec:Transmission}
First, we discuss the case where $N$ atoms are solely excited through the waveguide mode. This setting was theoretically studied in~\cite{mahmoodian_strongly_2018-1} and~\cite{kusmierek_higher-order_2023} based on a scattering matrix formalism and higher order cumulant expansion, respectively, with a discussion of the resonant excitation case ($\Delta =0$). Here, we extend this discussion to off-resonant excitation, which has been experimentally studied and theoretically analyzed~\cite{cordier_tailoring_2023} using the model presented here. 

The coherent amplitude of the output given by Eq.~\eqref{eq_alpha_wvd}. To evaluate it, we observe that the $n$-th atom is driven by the coherent amplitude $\alpha_{n}  = \alpha_{n-1} +  \alpha_\mathrm{sc,n-1} = t_\Delta \alpha_{n-1} = t_\Delta^{n-1} \alpha_\mathrm{in}$. One finds that the coherent component of the two-photon wavefunction decays exponentially according to Beer-Lambert's law:
\begin{equation}
    \psi_\mathrm{coh} = \alpha_\mathrm{out}^2  =  t_\Delta^{2N} \alpha_\mathrm{in}^2.
\end{equation}

In order to compute the incoherent component of the output according to  Eq.~\eqref{Phi_N_delta} in this configuration, we now have to calculate the local Rabi frequencies, which are given by:
\begin{equation}
\Omega_{n}^\mathrm{wg} = 2\sqrt{\beta \Gamma} \alpha_{n} = 2\sqrt{\beta \Gamma} t^{n-1}_\Delta \alpha_\mathrm{in} = t_\Delta^{n-1} \Omega_\mathrm{in}^\mathrm{wg}~.
\end{equation}
Inserting this expression into the incoherent two-photon component (see Eq.~\eqref{eq:phi}) emitted by the $n$-th emitter into the waveguide, we obtain:
\begin{align}
\psi^{(n)}_\mathrm{incoh}(\omega)&= -\frac{2 (t^{n-1}_\Delta\Omega^\mathrm{wg}_\mathrm{in})^2}{ \beta \Gammatot} g_\Delta g_{\Delta+\omega} g_{\Delta-\omega}~.
\end{align}

Summing up the incoherent contribution of all emitters at the waveguide output (see Eq. ~\ref{Phi_N_delta}), the incoherent contribution of all emitters, we obtain the incoherent component of the output two-photon wavefunction:
\begin{align}
    &\psi_\mathrm{incoh}(\omega) =  (\Omega_\mathrm{in}^\mathrm{wg})^2 \sum_{n = 1}^N t_\Delta^{2(n-1)}  \tilde{\psi}_\mathrm{incoh}(\omega) \left(t_{\Delta + \omega} t_{\Delta - \omega}\right)^{N-n} \nonumber\\
    &=(\Omega_\mathrm{in}^\mathrm{wg})^2 \, \tilde{\psi}_\mathrm{incoh}(\omega) \times \begin{cases}  \frac{t_\Delta^{2N}  - \left[t_{\Delta + \omega} t_{\Delta - \omega}\right]^{N}}{ t_\Delta^2  - t_{\Delta + \omega} t_{\Delta - \omega}} &\omega  \neq 0
\\N t_\Delta^{2(N-1)} &\omega =0
\end{cases}
\label{Phi_N_delta_trans}   
\end{align}
where in the last step we computed the geometric sum explicitly and we introduce the normalized incoherent two-photon component
\begin{equation}
    \tilde{\psi}_\mathrm{incoh}(\omega) = \psi^{(n)}_\mathrm{incoh}(\omega)/\left(\Omega_n^\mathrm{wg}\right)^2~.
\end{equation}

As evident from Eq.~\ref{Phi_N_delta_trans}, the incoherent component of the output two-photon wavefunction  $\psi_\mathrm{incoh}(\omega)$ results from an interplay of (i) attenuation of coherent light along the ensemble, (ii) incoherent scattering at each atom and (iii) absorption of the incoherent photons by the following atoms, as sketched in Fig.~\ref{fig_sketch_N_atoms}.a). From the total two-photon wavefunction at the output, $\psi(\tau)$ , we obtain the second-order correlation function $g^{(2)}(\tau)$:
\begin{equation}
    g^{(2)}(\tau) = \frac{| \alpha_\mathrm{in}^2 t_\Delta^{2N}   + \psi_\mathrm{incoh}(\tau)|^2}{ |\alpha_\mathrm{in}t_\Delta^N|^{4} }~.
    \label{eq:g2photontransport}
\end{equation} 
Here, as $\psi_\mathrm{incoh} \propto \alpha_\mathrm{in}^2$, $g^{(2)}(\tau)$ is independent of the input power. This is a consequence of the truncation to the two-photon number state used here~\footnote{Our theory is developed to leading order in $|\Omega|/\Gammatot$ as in~\cite{mahmoodian_strongly_2018-1}. Correction to our model are on the order $P_\mathrm{in}\propto |\Omega|^2$ for $g^{2}(\tau)$, and $P_\mathrm{}^2$ for the squeezing spectrum}.  

It is noteworthy that equation~\eqref{Phi_N_delta_trans} is valid for arbitrary atom-laser detunings and thus goes beyond what has been discussed in~\cite{mahmoodian_strongly_2018-1}. In the following, we will see that off-resonant excitation leads to richer interference phenomena compared to the resonant case. This stems from the fact that, in the detuned case, $\psi_\mathrm{coh}$ and $\psi_\mathrm{incoh}(\tau)$ are complex quantities, while they are real in the resonant case. 

In this context, it is useful to define the phase difference between the incoherent and coherent component at zero time delay, $\Delta  \varphi = \mathrm{Arg}\lbrace \psi_\mathrm{incoh}(\tau=0)\rbrace -  \mathrm{Arg}\lbrace \psi_\mathrm{coh}\rbrace$. Depending on $\Delta\varphi$, the incoherent two-photon component will interfere with the coherent component either destructively (when $\Delta \varphi = \left(2m+1\right) \pi,\,m \in \mathbb{Z}$) or constructively (when $\Delta \varphi = 2m \pi,\,m \in \mathbb{Z}$). This is illustrated in Fig.~\ref{fig_sketch_N_atoms} b) where we show the evolution of both the coherent and incoherent two-photon amplitude in the complex plane as a function of the number of emitters for two different detunings. The coherent component still decays exponentially with the number of emitters, as in the case for resonant excitation. By comparison, the incoherent two-photon amplitude at zero time delay $\psi_\mathrm{incoh}(\tau=0)$ shows a more complex dynamics. Its magnitude first linearly increases due to a collective enhancement in forward scattering~\footnote{See Appendix~\ref{sec_waveguide_illum_approx_formulaes} and~\cite{hinney_unraveling_2021} for more details.}. For larger atom number, the magnitude of the two-photon component decreases sub-exponentially with $N$~\cite{mahmoodian_strongly_2018-1}. 

We note that, for larger detunings, even richer dynamics takes place. Due to the dispersive action of the following atoms, the amplitude of the incoherent two-photon component emitted by atoms at the beginning of the chain can destructively interfere with that emitted by atoms at the end of the chain. This can lead to a vanishing incoherent component $\psi_\mathrm{incoh}(\tau = 0)$ for a certain combinations of detuning and atom number (see Appendix~\ref{sec_waveguide_illum_approx_formulaes} and Fig.~\ref{fig_regimes} therein).

Turning now to Fig.~\ref{fig_sketch_N_atoms}.c), it is apparent that off-resonant the combined dynamics of $\psi_\mathrm{coh}$ and $\psi_\mathrm{incoh}(\tau=0)$  leads to an oscillatory behavior of $g^{(2)}(\tau=0)$ as a function of $N$, which stems from the $N$-dependence of $\Delta \varphi$, see panel b), leading to alternating constructive and destructive interference of the two components. Surprisingly, when scanning $N$ and $\Delta$, we find that several combinations lead to perfect destructive interference and, hence, to photon antibunching (see orange and green lines in panel c). Specifically, this occurs when the moduli of both components are equal, $|\psi_\mathrm{coh}| = |\psi_\mathrm{incoh}(\tau=0)|$, and their phase difference is $\Delta \varphi = (2m +1)\pi$, $m\in \mathbb{Z} $.
While the antibunching point on resonance ($\Delta=0$) has been predicted and observed in~\cite{mahmoodian_strongly_2018-1,prasad_correlating_2020}, perfect antibunching for off-resonant excitations is a new prediction. It has been recently observed experimentally based on the predictions of this model~\cite{cordier_tailoring_2023}.

\begin{figure}
    \centering
    \includegraphics[width=\linewidth]{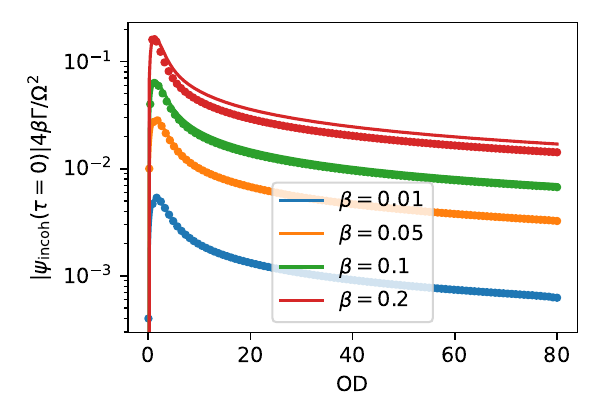}
    \caption{Comparison~\citet{mahmoodian_strongly_2018-1} (dots) and the model of this paper (solid lines) for different $\beta$ on the incoherent two-photon component $|\psi_\mathrm{incoh}(\tau=0)|$ as a function of optical depth $OD= 4\beta N$.  For small $\beta$ a very good agreement is achieved, even for very large ensembles. For larger $\beta$ deviations occur due to higher order scattering processes. }
    \label{fig_comp_Mahmoodian}
\end{figure}

We now compare our results to~\citet{mahmoodian_strongly_2018-1} where the two-photon wavefunction has been calculated using a scattering formalism together with the Bethe-Ansatz technique~\cite{shen_strongly_2007-2}. In contrast to our model, this approach yields exact solutions for any value of $\beta$. In Fig.~\ref{fig_comp_Mahmoodian} we compare the incoherent component $|\psi_\mathrm{incoh}(\tau=0)|$ calculated with the two approaches. Both predictions agree very well for $\beta \ll 1$. For larger $\beta$ deviations are expected as our model does not take into account higher order scattering processes. Note that, in contrast to~\cite{mahmoodian_strongly_2018-1}, our model provides a closed  expression for $ \psi_\mathrm{incoh}(\tau=0)$, which is given by the frequency integral of the analytical expression given in Eq.~\eqref{Phi_N_delta_trans}. Thus, its numerical complexity does not depend on $N$, contrary to the numerical calculation based on~\cite{mahmoodian_strongly_2018-1}. Consequently, predictions made with our model are numerically much less expensive for large $N$. 

\subsection{External illumination: Bragg angle }\label{sec_Bragg}

\begin{figure}[h]
\centering
\includegraphics[width=0.95\linewidth]{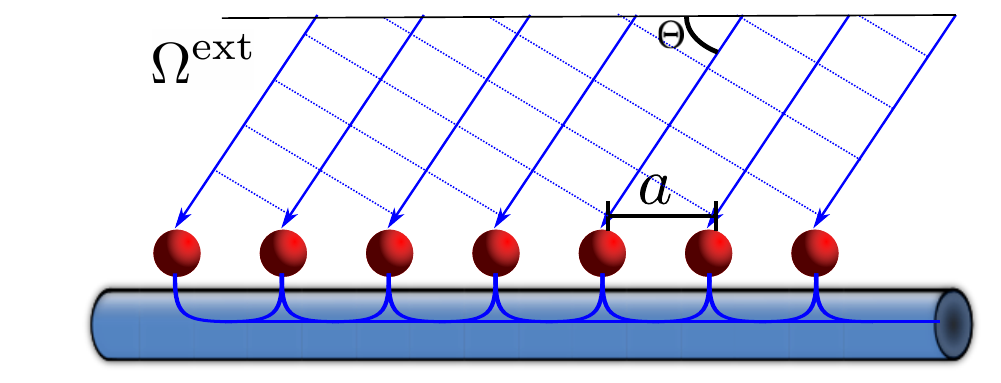}
\caption{Sketch of the external illumination by a resonant plane wave where $a$ is the distance between the atoms. Every atom is driven by field with a Rabi frequency of magnitude $|\Omega^\mathrm{ext} |$ under a Bragg angle $\Theta$. }
\label{fig_Bragg_sketch} 
\end{figure}

The next setting under study is an array of atoms externally illuminated by a resonant plane wave as sketched in Fig.~\ref{fig_Bragg_sketch}. 
Let us first consider an illumination under the Bragg angle $\Theta$ such that the partial amplitudes scattered by all $N$ emitters interferes constructively in the waveguide. This Bragg condition is fulfilled when:
\begin{equation}
    \cos \Theta=\frac{m \lambda}{a}-\frac{\lambda}{\lambda_{\mathrm{f}}}\label{eq_Bragg_cond}
\end{equation}
where $\lambda_f = \lambda/n_\mathrm{eff}$ is the wavelength inside the waveguide, $n_\mathrm{eff}$ is the effective refractive index of the guided mode, and $a$ the lattice spacing and $m$ the Bragg order.
This situation resembles the situation studied in~\cite{jones_collectively_2020,olmos_interaction_2020}. We will neglect free space interaction between the atoms such that we only consider Bragg conditions fulfilling $a\gtrsim \lambda$. We note that possible voids in the atomic array would not affect our predictions.

For a plane wave, each emitter is subject to an identical external driving field with a Rabi frequency of magnitude $|\Omega^\mathrm{ext}|$. The first emitter scatters a coherent amplitude $\alpha_{1}^\mathrm{sc}$ into the waveguide which constructively interferes with the scattered amplitude from the second emitter according to the Bragg condition of Eq.~\eqref{eq_Bragg_cond}. After $N$ emitters, and taking into account absorption, the output two-photon coherent amplitude inside the waveguide is 
\begin{align}
  \alpha_\mathrm{out} &= \sum_{n=1}^N  t_{0}^{N-n} \alpha_{1}^\mathrm{sc} =  \frac{1-t_0^N}{1-t_0} \alpha_{1}^\mathrm{sc}  = \frac{1-t_0^N}{2\beta} \alpha_{1}^\mathrm{sc}  .
\end{align}
The atoms are now driven by the field that results from the interference between the external drive and the fiber guided field. At resonance and under Bragg condition, these two fields are in phase opposition at each atom. Specifically, the Rabi frequency seen by the $n$-th emitter is given by ${\Omega_n} = t_0^{n-1} {\Omega^\mathrm{ext}} $ which interestingly decreases with the atom number. This exponential decrease is consistent with findings in~\cite{olmos_bragg_2021} on Bragg-scattering from optically dense ensembles. From the driving amplitude $\Omega_n$ we obtain the incoherent two-photon amplitude 
\begin{equation}
    \psi_\mathrm{incoh}(\omega) =  (\Omega^\mathrm{ext})^2 \sum_{n = 1}^N t_0^{2(n-1)}  \tilde{\psi}_\mathrm{incoh}(\omega) \left|t_{ \omega}\right|^{2(N-n)},
    \label{eq_phi_N_Bragg}
\end{equation}
using $t_{-\omega} = t_\omega^*$.
Interestingly, the expression of $\psi_\mathrm{incoh}(\omega)$ in Eq.~\eqref{eq_phi_N_Bragg} is the same as for an illumination through the waveguide in Eq.~\eqref{Phi_N_delta_trans} for $\Delta =0$, i.e. both incoherent two-photon amplitudes are equal if the external Bragg field drives the first atom with the same strength as in the case where we probe the system through the waveguide. The coherent two-photon amplitude, however, differs strongly, and for external illumination we obtain:
\begin{equation}
    \psi_\mathrm{coh} =  \frac{\left(1- t_0^N \right)^2}{4\beta^2} \left(\alpha_{1}^\mathrm{sc}\right)^2
\end{equation}

\begin{figure}[h]
\centering
\includegraphics[width=0.9\linewidth]{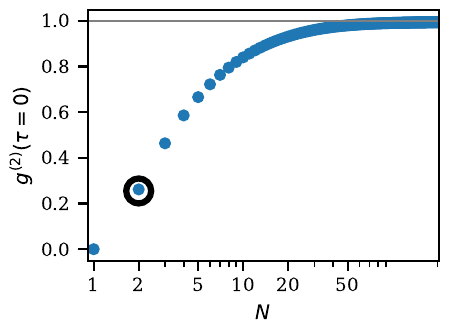}
\caption{Second-order correlation function at zero time-delay $g^{(2)}(0)$ under the Bragg condition as a function of $N$ for $\beta = 0.01$. 
As in the excitation scheme with a single emitter from the side (Sec.~\ref{sec_single_emitter_external}), the point of perfect anti-bunching appears at $N=1$. For larger $N$, $g^{(2)}(0)$ approaches 1. Interestingly, for $N=2$ an antibunching of $g^{(2)}(0)$ of 1/4 is predicted (black circle).}
\label{fig_Bragg_g2N} 
\end{figure}
 This leads to the normalized second-order correlation function
\begin{equation}
    g^{(2)}(\tau) = \frac{|\left(\alpha_{1}^\mathrm{sc}\right)^2 (1-t_0^N )^2 + 4\beta^2 \psi_\mathrm{incoh}(\tau)|^2}{|\alpha_{1}^\mathrm{sc}(1-t_0^N )|^4}.
\end{equation}
As expected, for $N=1$, one gets antibunching $g^{(2)}{(0)}=0$ (fluorescence from a single emitter) and for large $N$ the output light approaches $g^{(2)}{(0)}=1$ as shown in Fig.~\ref{fig_Bragg_g2N}.

It is interesting to note that for $N=2$ we obtain $g^{(2)}(0) \simeq 1/4$. This is in contrast to the criteria of $g^{(2)}(0) \leq 1/2$ that is commonly used to verify the presence of a single quantum emitter. 
A $g^{(2)}(0)$ smaller than 1/2, in our case, is due to the constructive interference of the incoherent photons under Bragg illumination from different emitters. In contrast, if we consider $N$ emitters with shot-to-shot fluctuations of their relative distances, one recovers the common criteria $g^{(2)}(0) = 1 - 1/N$ (see Appendix~\ref{sec_random_dist}). 
We note that $g^{(2)}(0)<1/2$ from two coherently emitters has been  predicted in the literature~\cite{mandel_photon_1983,skornia_nonclassical_2001}.

\subsection{External illumination: anti-Bragg}
\label{sec_anti_Bragg}

\begin{figure}[tb]
    \centering
    \includegraphics[width=0.95\linewidth]{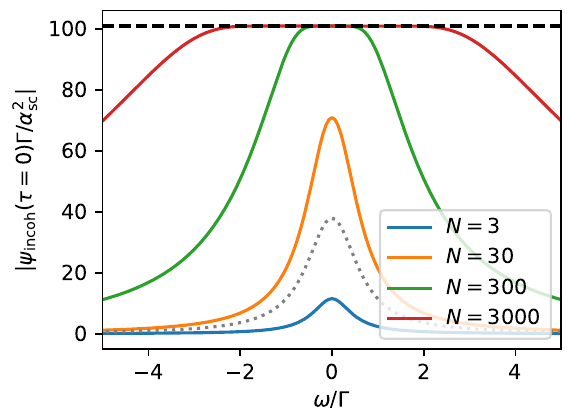}
    \caption{Spectrum of the incoherent component $\left|\psi_\mathrm{incoh}(\omega)\Gammatot/\alpha_\mathrm{sc}^2\right|$ for atoms illuminated under the anti-Bragg angle condition and $\Delta=0$. The curves are obtained for $\beta = 0.01$ and different atom numbers. For small $N$, a coherent built-up takes place, while for large $N$ the spectrum saturates to $\alpha^2_\mathrm{sc}/(\beta(1-\beta)$ as shown by the black dashed line. For comparison, the dotted gray line shows $\left|\psi_\mathrm{incoh}(\omega)\Gammatot/\alpha_\mathrm{sc}^2\right|$ for illumination through a waveguide~\cite{hinney_unraveling_2021} (or equivalently in Bragg configuration) for the same $\beta$ and the atom number that maximizes $\left|\psi_\mathrm{incoh}(\omega)\right|$ at $\omega =0$. }
    \label{fig_anti_Bragg}
\end{figure}

\begin{figure*}[btp]
\centering
\includegraphics[width=0.9\linewidth]{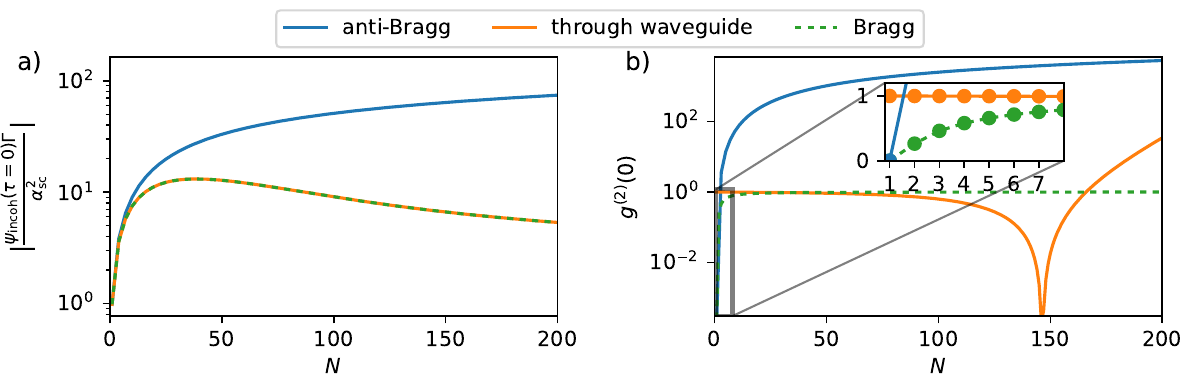}
\caption{a) Incoherent component of the two-photon wavefunction $|\psi_\mathrm{incoh}(\tau=0)|$ as a function of the number of emitters $N$. In contrast to the Bragg/transmission illumination $|\psi_\mathrm{incoh}(\tau=0)|$ grows much faster in the anti-Bragg configuration. For comparison, the two-photon field is normalized to the scattered coherent field $\alpha_\mathrm{sc}^2$ of the first emitter in both cases. b) The second-order correlation function $g^{(2)}(0)$ as a function of emitters $N$ in solid blue for the anti-Bragg configuration and for odd $N$. For comparison, $g^{(2)}(\tau=0)$ for the transmission configuration is shown in orange and for the Bragg configuration in green, where $g^{(2)}(0)$ quickly reaches $1$ after an initial antibunching for $N=1$ (shown in the inset zoom). Both figures are calculated for $\beta = 0.01$.}
\label{fig_anti_Bragg_int_squeezing_and_g2}
\end{figure*}

Now we look at the anti-Bragg configuration~\cite{poddubny_driven_2022}, i.e., the case where emitted coherent fields interfere destructively inside the waveguide, in contrast to the Bragg case. Again, we restrict our discussion to the resonant case and the anti-Bragg angle $\zeta$ is given by:
 \begin{equation}
    \cos \zeta=\left(m + \frac{1}{2}\right)\frac{\lambda}{a}-\frac{\lambda}{\lambda_{\mathrm{f}}}.
    \label{eq_anti_Bragg_cond}
\end{equation}
In this anti-Bragg configuration, the coherent amplitude inside the waveguide depends on whether the atom number is odd or even: 
\begin{equation}
    \psi_\mathrm{coh} = \begin{cases} \alpha^2_\mathrm{sc} &\text{for N odd},\\
    0   &\text{for N even.}
    \end{cases}
\end{equation}
Thus, in both cases, the scattering amplitude is close to zero for small coupling strength $\beta$. 
In the following, we will neglect the driving of the atoms via this small field $\alpha_\mathrm{sc}$~\footnote{This approximation is well fulfilled for $\beta \ll 1$.} and assume that all emitters are solely driven by the external drive $\Omega^\mathrm{ext}$.

The incoherent two-photon field thus becomes:
\begin{equation}
    \psi_\mathrm{incoh}(\omega) = \big({\Omega^\mathrm{ext}}\big)^2 \sum_{n = 1}^{N} \tilde{\psi}_\mathrm{incoh}(\omega) \left|t_{ \omega}\right|^{2(N-n)}
\end{equation}
this geometric sum can be simplified to
\begin{equation}
    \psi_\mathrm{incoh}(\omega) = \big({\Omega^\mathrm{ext}}\big)^2  \tilde{\psi}_\mathrm{incoh}(\omega) \frac{1- \left|t_{\omega}\right|^{2N}}{1- |t_{\omega}|^2 }.\label{phi_N_anti_Bragg_full} 
\end{equation}
In contrast to the above case studies, the driving Rabi frequency remains constant along the chain, which leads to a continuous build-up of the incoherent component with $N$ and different spectral properties of the incoherent component. In Fig.~\ref{fig_anti_Bragg} we show $|\psi_\mathrm{incoh}(\omega)|$. It is apparent that, for large enough $N$, such that $|t_{\omega=0}|^{N}\simeq 0$, $\psi_\mathrm{incoh}(\omega)$  exhibits a plateau around $\omega=0$. This plateau asymptotically reaches a value of $-\alpha_\mathrm{sc}^2\Gammatot/(\beta(1-\beta))$, as indicated by the black dashed line in Fig.~\ref{fig_anti_Bragg}.

The strong incoherent component $\psi_\mathrm{incoh}(\omega)$ can be directly linked to squeezed light via Eq.~\eqref{eq_squeezing}, and we obtain a maximal squeezing of 
\begin{equation}
    S_0(\omega) = -\frac{\big(\Omega^\mathrm{ext}\big)^2}{\Gamma^2 (2-2\beta)}~\label{eq_squeezing_anti_Bragg}
\end{equation}
 within the plateau and we note that Eq.~\eqref{eq_squeezing_anti_Bragg} is only valid for $\Omega^\mathrm{ext} \ll \Gamma$~\footnote{Specifically, it should not be concluded that perfect squeezing (i.e., $S_0(\omega) = -1/4$) can be achieved as our model loses validity when $\Omega_\mathrm{ext} \simeq \Gammatot$.}.

To quantify the strength of the atom's non-linearity, we integrate $\psi_\mathrm{incoh}(\omega)$ over all frequencies,  $\psi_\mathrm{incoh}(\tau=0) = \int_{-\infty}^{\infty} \frac{\mathrm{d}\omega}{2\pi} \psi_\mathrm{incoh}(\omega)$. This quantity can be experimentally inferred from the noise reduction in the amplitude quadrature~\cite{hinney_unraveling_2021}. 

To compare different configurations, we normalize $\psi_\mathrm{incoh}(\tau)$ to the scattering rate of the first emitter $\left(\alpha_{1}^\mathrm{sc}\right)^2$. With this normalization, $\psi_\mathrm{incoh}(\tau)$ becomes independent of the input power and the three different configurations are compared in
Fig.~\ref{fig_anti_Bragg_int_squeezing_and_g2}. 
From Fig.~\ref{fig_anti_Bragg_int_squeezing_and_g2} a) it is apparent that for the same number of atoms $|\psi_\mathrm{incoh}(\tau=0)|$ reaches much larger values in the anti-Bragg configuration than under illumination through the waveguide (or, equivalently, for the Bragg-configuration).  This larger amount of incoherent photon pairs is due to the fact that the driving strength along the chain does not decrease, in contrast to the exponentially damped driving in both Bragg illumination and illumination through the waveguide.

Let us finally discuss the normalized second-order correlation function $g^{(2)}(\tau)$ in the anti-Bragg configuration, see Fig.~\ref{fig_anti_Bragg_int_squeezing_and_g2} b). In  our approximation of Eq~\eqref{eq:g2}, it diverges for every even number of atoms, since the coherent power $P_\mathrm{coh} \propto |\alpha_\mathrm{out}|^2 $ is zero~\footnote{Note that this is not a limitation of our model, as this divergence in $g^{(2)}(0)$ strictly only occurs for vanishing input power. For finite input power high-order corrections will lead to finite photon bunching.}. 
For odd atom numbers, one obtains:
\begin{equation}
g^{(2)}(\tau) = | 1  -  \frac{\psi_\mathrm{incoh}(\tau)}{\alpha_\mathrm{sc}^2}|^2.\label{eq_g2_anti_Bragg} 
\end{equation} 
While the coherent two-photon component is independent of $N$ with a value of $\alpha^2_\mathrm{sc}$, the incoherent two-photon component $\psi_\mathrm{incoh}(\tau =0)$ grows much faster than in the two other configurations. This leads to strong bunching already for much smaller atom numbers that when illuminating through the waveguide, see Fig.~\ref{fig_anti_Bragg_int_squeezing_and_g2} b). Thus, the anti-Bragg configuration is promising for solid state emitters, where obtaining large ensembles of identical emitters is challenging. For example, already for two emitters, perfect bunching (i.e., diverging $g^{(2)}(\tau=0)$) is predicted~\cite{wolf_light_2020}. Note that, the underlying destructive interference could be even obtained for randomly spaced emitters with inhomogeneous transition strengths and frequencies. This only requires individual addressability including the ability to accordingly adapt the local phases and amplitudes of the external drive, e.g., with a spatial light modulator (SLM).
The increased yield of correlated photon pairs could then be of practical use for narrowband photon pair sources, or heralded single photon sources.

\subsection{Combined illumination through waveguide and external}\label{sec_combined}
\begin{figure}[h]
    \centering
    \includegraphics[width=\linewidth]{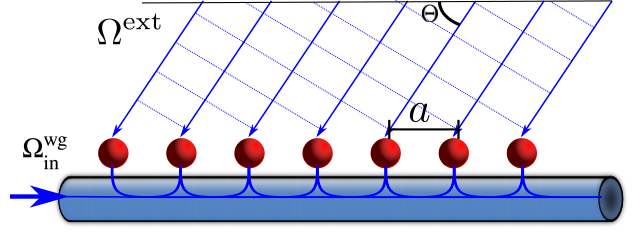}
    \caption{Combined illumination through the waveguide and externally under the Bragg angle $\Theta$. This configuration effectively increases the coupling strength $\beta'$, leading to strong non-linear effects with fewer emitters.
    }
    \label{fig_Bragg_and_transmission}
\end{figure}
In this last section, we study the case where many atoms are driven simultaneously via the waveguide and an external field as sketched in Fig.~\ref{fig_Bragg_and_transmission} with both fields in phase at the atoms. As discussed in Sec.~\ref{sec_single_combined} for a single atom and $\beta \ll 1$, the light in the fiber effectively behave as the atoms were solely illuminated through the waveguide with a rescaled coupling constant $\beta'  = \beta \left( 1 + \Omega^\mathrm{ext}/\Omega^\mathrm{wg}\right) $.

Then, this concept of an effectively increased coupling constant greatly simplifies the problem. Specifically, the output coherent photon component for $N$ emitters and resonant drive ($\Delta=0$) is given by
\begin{equation}
    \alpha_\mathrm{out} = \alpha_\mathrm{in} \prod_{n=1}^N t'(n)
\end{equation} 
where the individual amplitude transmission coefficients are given by $t'(n) = 1 - 2\beta'(n)$. Here, the coupling strength of the n-th emitter is given by
\begin{equation}
    \beta'(n) / \beta =  1+ \Omega^\mathrm{ext}/( \Omega_n^\mathrm{wg} t^{n-1}_0) .
\end{equation}

For the incoherent component, the creation process behaves as if each coupling constant was $\beta'(n)$. However, the transimission of the incoherent photon pairs $|t_{\omega}|^{2(N-n)}$ continues to scale with the bare value of $\beta$ . This leads to:
\begin{equation}
    \frac{\psi_\mathrm{incoh}(\omega)}{\alpha^2_\mathrm{in}} =  -  \sum_{n = 1}^N \left(\prod_{i=1}^{n-1} t'(i) \right)\frac{2 g'(n) \, |g'_{\omega}(n)|^2 }{\beta'(n) \Gamma}  \left|t_{\omega} \right|^{2(N-n)}
\end{equation}
 where $|\alpha_\mathrm{in}|^2$ is the power sent through the waveguide, and $g_\omega'(n)= \frac{2\beta'(n)}{1 - 2i\omega/\Gammatot} $ the position dependent photon generation coefficient. In analogy to $t'(n)$, we omit the subscript for $g'(n)$, as here we assumed $\Delta=0$. 

\begin{figure}[h]
    \centering
    \includegraphics[width=\linewidth]{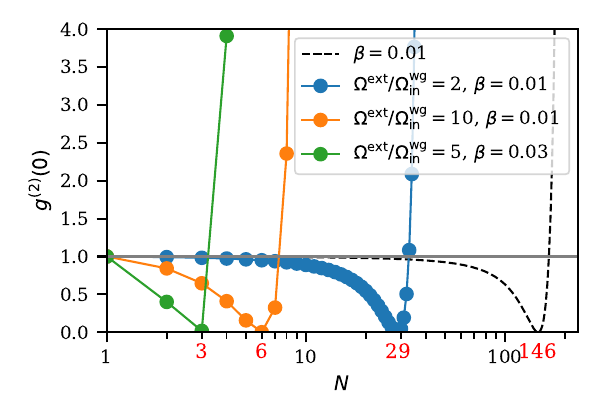}
    \caption{The second-order correlation function $g^{(2)}(0)$ as function of the numbers of emitters in combined illumination under the Bragg angle and through the waveguide for $\Delta=0$ for three exemplary parameter sets. For comparison, the black dashed line shows the case of pure waveguide illumination with a bare coupling constant $\beta=0.01$. By adding the external Bragg beam with a field strength $\Omega^\mathrm{ext}/ \Omega^\mathrm{wg}_\mathrm{in}= 2$ at the first emitter, the collective non-linear interaction of the ensemble is enhanced such that for $N=29$ a point of perfect antibunching is reached (orange line). For $\Omega^\mathrm{ext}/ \Omega^\mathrm{wg}_\mathrm{in}= 10$ this point is already reached with 6 atoms. When increasing the coupling strength this effect becomes more pronounced, and with $\beta =0.03$ perfect antibunching can be reached with three emitters and  $\Omega^\mathrm{ext}/ \Omega^\mathrm{wg}_\mathrm{in}= 5$. The solid lines are guides to the eyes.}
    \label{fig_Bragg_and_transmission_N}
\end{figure}

Let us discuss some concrete examples. For a typical value of $\beta = 0.01$, as obtained in cold-atom nanofiber setups, introducing an external Bragg field with an amplitude twice that of the nanofiber-guided light $\left( {\Omega^\mathrm{ext}}/{\Omega_n^\mathrm{wg}} = 2 \right)$ enhances the effective coupling strength of the first atom to $\beta' = 3\beta$, and to even larger values for the following atoms (see Appendix~\ref{app_combined}). This drastically changes the behavior of the ensemble. Particularly, the number of emitters required for perfect antibunching in transmission~\cite{prasad_correlating_2020} reduces from 146 to 29 as shown in Fig.~\ref{fig_Bragg_and_transmission_N}. For moderately larger $\beta = 0.03$, as readily obtained in solid state systems, and $\Omega^\mathrm{ext}/ \Omega^\mathrm{wg}_\mathrm{in}= 10$, the number of emitters required for perfect antibunching even reduces to three. 

These last two examples illustrate how an external drive can help reducing the number of identical quantum emitters required to obtain (anti-)bunched light from an ensemble.
Such a reduction is especially interesting for implementations using solid states systems where obtaining large number of identical emitters is challenging. Moreover, as discussed in Sec.~\ref{sec_anti_Bragg}, the required phase relationship between the waveguide light and the external drive could even be met for randomly placed emitters by adapting the local phases, e.g., with a spatial light modulator (SLM). Moreover,  a variation of the bare coupling strength between the emitters can also be compensated with an SLM, or could be readily taken into account in our model. Thus, based on this approach, strong antibunching or bunching, can be reached in many experimental situations and physical platforms.

\section{Conclusion \& Summary}
In summary, we have presented a theoretical model to efficiently calculate the nonlinear 
response of an ensemble of $N$ quantum emitters weakly coupled ($\beta \ll 1$) to a single propagating optical mode in the low saturation regime ($\Omega \ll \Gamma$).
Using this model, we derive simple analytic expressions for the incoherent component of the two-photon wavefunction, $\psi_\mathrm{incoh}(\omega)$, for numbers of emitters for which alternative state-of-the-art methods~\cite{mahmoodian_strongly_2018-1,kusmierek_higher-order_2023} are numerically costly. 
This allows us to infer both the squeezing spectrum and the second-order correlation function.
We show how large ensembles, as well as countable numbers of identical weakly-coupled emitters can produce quantum states of light with sub/super Poissonian statistics and squeezing. While this is technologically achievable with cold atoms coupled to nanofibers as demonstrated in \cite{prasad_correlating_2020,hinney_unraveling_2021,cordier_tailoring_2023}, obtaining large ensemble of identical emitters is challenging with solid state systems. To circumvent this limitation, we present two techniques that  allow one to arbitrarily reduce the number of emitters required to observe these strong quantum effects. 

It is an open question how the discussed effects extend to larger saturation. Recently, on-resonant illumination through the waveguide was studied also for larger saturation parameters~\cite{kusmierek_higher-order_2023} but remains unexplored in the off-resonant case, which is also covered by our model. 
Furthermore, our model could be extended such that it also yields reliable predictions for larger values of $\beta$. In this regime, interesting effects such as photon number sorting~\cite{mahmoodian_dynamics_2020-1}, topologically protected edge states~\cite{mcdonnell_subradiant_2022,svendsen_topological_2024}, and many-body-localized phases~\cite{fayard_many-body_2021} have been predicted. Here, we expect our model to yield physical insights in the case of large ensembles where other approaches exclusively offer numerical predictions.

\begin{acknowledgments}
We gratefully thank S. Mahmoodian, K. Hammerer and A. S{\o}rensen for discussions and feedback.  M.S. acknowledges funding Italian Ministry of University and Research, in the framework of the National Recovery and Resilience Plan (NRRP) for funding project MSCA\_0000048. M. C. and M. S. acknowledge support from the European Commission (Marie Skłodowska-Curie IF Grant No.~101029304 and IF Grant No.~896957). We acknowledge funding by the Alexander von Humboldt Foundation in the framework of the Alexander von Humboldt Professorship endowed by the Federal Ministry of Education and Research, as well as funding by the European Commission under the project DAALI (No.~899275), and by the Einstein Foundation (Einstein Research Unit on Quantum Devices). 
\end{acknowledgments}

\appendix

\section{Unidirectional coupling}\label{sec_SM_unidirectional} 
In this article, we assume a perfectly chiral system where all emitters scatter only in forward direction of the waveguide.
This assumption is motivated by several factors: First, such chiral scattering exits naturally in setups based on cold atoms coupled to nanofibers~\cite{sayrin_nanophotonic_2015}. Second, unidirectional coupling drastically reduces the complexity of the calculation, and third, for \textit{illumination through the waveguide} we expect only small differences between chiral and symmetric coupling~\cite{mahmoodian_strongly_2018-1}. The same is expected to hold for the \textit{Bragg configuration} and \textit{anti-Bragg configuration}.

\section{Illumination through the waveguide: approximate formulas for $\psi_\mathrm{incoh}(\tau)$ and $g^{(2)}(\tau)$ for resonant excitation $\Delta=0$}\label{sec_waveguide_illum_approx_formulaes}
\begin{figure*}[htbp]
\centering
\includegraphics[width=\linewidth]{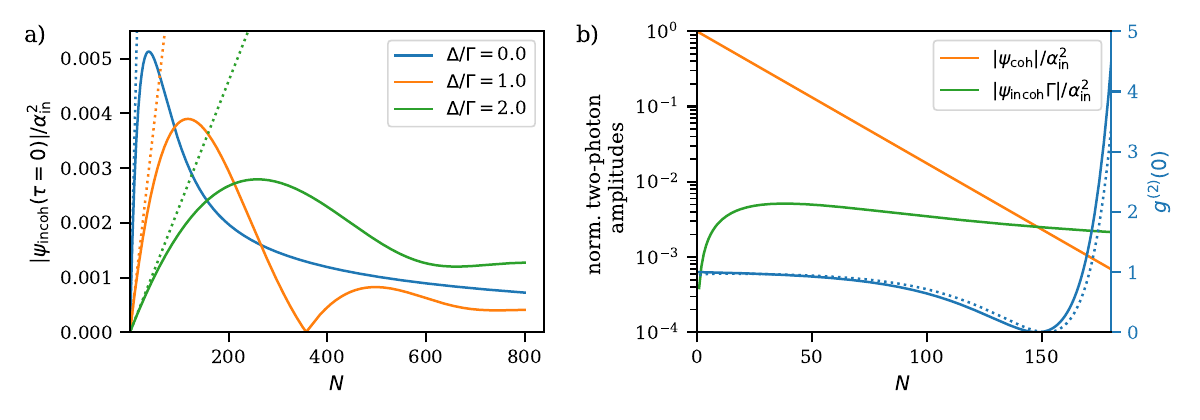}
\caption{a) The incoherent component $\psi_\mathrm{incoh}(\tau=0) $ for $\beta= 0.01$ and different detuning $\Delta$ as a function of the emitter $N$. In the off-resonant case $\psi_\mathrm{incoh}(\tau=0)$ shows a complicated dynamic. For small number of emitters (low OD-regime), it first increases linearly with $N$ due to the coherent enhancement. For larger $N$ a more complicated picture emerges due to interferences between the light scattered by different emitters that can even lead to a cancellation of $\psi_\mathrm{incoh}(\tau=0)$. The dotted lines show the low OD approximation of coherent build-up in forward scattering. b) Evolution of the incoherent and coherent component at zero delay, as well as the corresponding $g^{(2)}(\tau=0)$ as a function of $N$ for the same $\beta$ and $\Delta=0$. At $N=150$ the point of destructive interference occurs and for larger $N$ strong bunching can be observed~\cite{mahmoodian_strongly_2018-1,prasad_correlating_2020}. The dotted line is the further approximation of Eq.~\ref{eq_g20_approx} with good agreement for small $\beta$.}
\label{fig_regimes}
\end{figure*}

In the configuration where the laser illuminates the emitters through the waveguide (Sec.~\ref{sec:Transmission}), the magnitude of the coherent component $\psi_\mathrm{coh}$  decays exponentially with $N$ for any $\Delta$, while the incoherent component $\psi_\mathrm{incoh}$ exhibits a more complex dynamics, as discussed in the following.

\paragraph{Off-resonant}

As shown in Fig.~\ref{fig_regimes} a), for a detuned excitation field  $\psi_\mathrm{incoh}(\tau =0)$ shows a complicated dynamics as a function of $N$ and $\Delta$. Since the phase of the generated incoherent component changes along the waveguide, even destructive interference between the newly generated incoherent component and the pre-existing one can occur, as shown by the orange line in Fig.~\ref{fig_regimes} a).

Still, a simple picture can be provided by introducing the optical depth $OD=-2N \log (|t_\Delta|)$ to distinguish the small and large OD regime. For small $OD$ the absorption can be neglected, i.e. $t_\omega^N\simeq 1$ for all $\omega$.
This implies that both the laser attenuation $t_\Delta^{2(n-1)} $ and the absorption of the incoherent component $ (t_{\Delta + \omega} t_{\Delta - \omega})^{N-n} $ can be neglected, and Eq.~\eqref{Phi_N_delta_trans} simplifies to
\begin{equation}
    \psi_\mathrm{incoh}(\omega) \underset{OD \ll 1}{=}  N \psi^{(1)}_\mathrm{incoh}(\omega)\label{eq_phi_N_low_OD}
\end{equation}
where $\psi^{(1)}_\mathrm{incoh}(\omega)$ is the incoherent component scattered by the first emitter.  
Eq.~\eqref{eq_phi_N_low_OD} reflects the coherent build-up of the incoherent component in forward direction, which is shown in Fig.~\ref{fig_regimes} a) by the dotted lines.

\paragraph{Resonant excitation}
It is instructive to consider the case of resonant excitation as in~\cite{prasad_correlating_2020}  which is sketched in Fig.~\ref{fig_regimes} b) for $\tau=0$. The incoherent component (orange) first increases linearly with $N$ in the low OD-regime. For larger OD, absorption cannot be neglected anymore, and $\psi_\mathrm{incoh}(\tau=0)$ decreases sub-exponentially. Due to this sub-exponenital decay the coherent component reaches at some point the magnitude of the exponentially decaying coherent component. At the point where both amplitudes are equal a destructive interference in the two-photon wavefunction $\psi(\tau=0)$ takes places which leads to antibunching with $g^{(2)}(0)=0$.

Mathematically, one can obtain insight by the following approximations: For small $\beta$, the discretization of the atoms can be neglected and $t_\omega^{N}\simeq e^{-\frac{OD}{2-4i\omega/\Gammatot}} $. For a resonant drive ($\Delta=0$) and in first order in $\beta$, Eq.~\eqref{Phi_N_delta_trans} simplifies to:
\begin{equation}
    \psi_\mathrm{incoh}(\omega) \underset{OD \gg 1}{=} \frac{\alpha_\mathrm{in}^2 \beta \Gammatot}{\omega^2}\left[e^{-OD \frac{1}{1+4{\omega^2}/{\Gammatot^2}}}- e^{-OD} \right] \,\mathrm{\forall \omega \neq 0}. \label{eq_phi_N_approx}
\end{equation}
For a larger number of emitters, Eq.~\eqref{eq_phi_N_approx} predicts sidebands which develop due to the absorption of the incoherent photon pairs from the ensemble of emitters and which have been observed in~\cite{hinney_unraveling_2021}. Note that Eq.~\eqref{eq_phi_N_approx} has a divergence for $\omega = 0$.  To simplify Eq.~\eqref{eq_phi_N_approx} one can introduce the approximation $1+4\omega^2/\Gammatot^2\simeq \omega^2\Gammatot^2$ in the exponent. This is a good approximation in the large OD regime where most of the incoherent photons are in the sidebands and around $\omega/\Gammatot \simeq 1$ the weight of $\psi_\mathrm{incoh}(\omega)$ is small. From this we can calculate:
 \begin{equation}
      \psi_\mathrm{incoh}(\tau= 0) = \frac{1}{2\pi} \int_{-\infty}^\infty  \mathrm{d}\omega \psi_\mathrm{incoh}(\omega)  = -\alpha_\mathrm{in}^2\sqrt{\frac{\beta}{4\pi N}}.
 \end{equation}
and it follows
\begin{equation}
    g^{(2)}(\tau=0) \approx \left[ 1 -  e^{4\beta N} \sqrt{\frac{\beta}{4 \pi N}} \right]^2 \label{eq_g20_approx}
\end{equation}
Note that, Eq.~\eqref{eq_g20_approx} represents a very good approximation for $g^{(2)}(0)$ even for a small OD, as shown in Fig.~\ref{fig_regimes}.  This is due to the fact that in the low $OD$-limit, $g^{(2)}(0)$ is exponentially dominated by the coherent component $\alpha_\mathrm{in}t_\Delta^{2N}$ and the error on $\psi_\mathrm{incoh}(\tau=0)$ has a negligible influence. A similar expression was found in~\cite{sheremet_waveguide_2023} based on a Green's function approach.

\section{External excitation with $N$-emitters at random distances}\label{sec_random_dist}
In this section we investigate  the situation of $N$ emitters which are located at random distances and calculate the second order correlation function $g^{(2)}(\tau)$. The light which arrives at atom $i$ is characterized by a phase $e^{i\varphi_i}$. Furthermore, for simplicity we restrict our analysis to resonant excitation ($\Delta = 0$)~\footnote{The same arguments holds for $\Delta \neq 0$.} and the low OD regime. In this regime, two effects can be neglected: absorption of the coherent component inside the waveguide, and absorption of the incoherent component. Then, the power at the waveguide output is given by
\begin{equation}
P_{\mathrm{wg}} = E\lbrace|\sum_{i=1}^N \alpha_\mathrm{sc} e^{i\varphi_i}|^2 \rbrace= N |\alpha_\mathrm{sc}|^2 
\end{equation}
where $E\lbrace\ldots\rbrace$ denotes averaging over the atomic position/phases $\varphi_i$. Similarly, the second-order correlation function $G^2(\tau)$ averaged over the atomic distances is
\begin{equation}
G^2(\tau) = E\left\lbrace  \left| \left( \sum_{i=1}^N \alpha_\mathrm{sc} e^{i\varphi_i}\right)^2   + \sum_{i=1}^N e^{i2\varphi_i} \alpha_\mathrm{sc}^2 e^{-\Gamma|\tau|}\right|^2\right\rbrace
\end{equation} 
For simplicity, let us focus on $\tau = 0$ and $\Delta=0$, and we obtain:
\begin{equation}
g^2(\tau = 0) = \frac{N^2 - N}{N^2} = 1 - \frac{1}{N}.
\end{equation}
This corresponds to the standard results for $g^{(2)}(0)$ of a Fock state with photon number $n$ and $n=N$. This is in contrast to our result in the \textit{Bragg-configuration} where we obtain approximately $g^{(2)}(0) \simeq |1-\frac{1}{N}|^2$ (neglecting absorption).

\section{Combined illumination through waveguide and external, discussion of $\beta'(n)$.}\label{app_combined}
\begin{figure}[h!]
    \centering
    \includegraphics[width=0.95\linewidth]{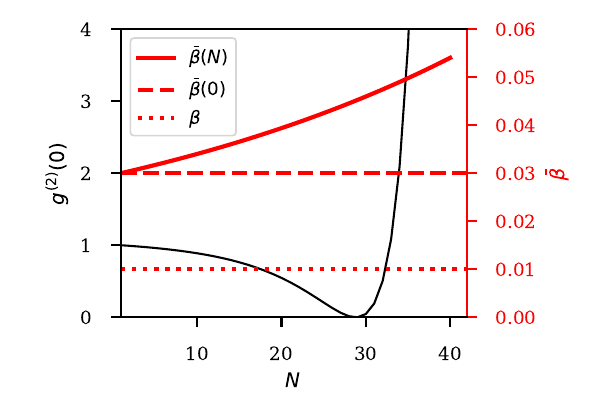}
    \caption{The left y-axis shows $g^{(2)}(0)$ for a combined Bragg and waveguide illumination for the same parameters as the blue line in Fig.~\ref{fig_Bragg_and_transmission}. The right y-axis shows the increase of the effective coupling constant ${\beta'}$ along the waveguide.}
    \label{fig_combined_beta}
\end{figure}
In principle, as in~\cite{goncalves_unconventional_2021-1}, the effective coupling constant ${\beta'(n)}$ is not bound by $1$ and $\beta'(n)$ diverges with increasing $N$. However, we can numerically show that for reasonable parameters $\beta'(n)$ does not exceed $1$ around the interesting point of antibunching, as shown in Fig.~\ref{fig_combined_beta} and our model stays valid with $\beta'(n) \ll 1$.


\bibliography{simple_model}

\end{document}